\documentclass[12pt,a4paper]{article}   
\usepackage{latexsym,graphicx,amssymb,amsmath} 
\usepackage{setspace,bm} 
\usepackage[numbers,sort&compress]{natbib}
\usepackage{changes}

\newcommand{\bea}{\begin{eqnarray}}
\newcommand{\eea}{\end{eqnarray}}
\newcommand{\be}{\begin{equation}}
\newcommand{\ee}{\end{equation}}

\newcommand{\rt}[1]{{}}
\newlength{\szovszel}
\newlength{\slashszel}

\begin{document}

\title{Radiation backreaction in axion electrodynamics}
\author{A. Patk\'os
\\
Institute of Physics, E{\"o}tv{\"o}s University\\
H-1117 P\'azm\'any P\'eter s\'et\'any 1/A, Budapest, Hungary\\
}
\vfill
\maketitle
\begin{abstract}
Energy-momentum conservation of classical axion-electrodynamics is carefully analyzed in the Hamiltonian formulation of the theory. The term responsible for the energy transfer between the electromagnetic and the axion sectors is identified. As a special application  the axion-to-light Primakoff-process in the background of a static magnetic field is worked out and the radiative self-damping of the axion oscillations is characterized quantitatively.  The damping time turns out comparable to the age of the Universe in the preferred axion mass range.
\end{abstract}

\section*{Homage to Professor Zhengdao Li}

50 years ago Professor Li visited Hungary among the distinguished speakers at the First International Conference on Neutrino Physics and Astrophysics held in Balatonf\"ured from 11th to 17th June 1972 (see Fig.1). His friendly but very sharp dispute with R. Feynman on the interpretation of the scaling phenomenon then freshly discovered in deep inelastic electron-proton scattering left lifelong impression in the audience, including also the present author. The concurrence of the Friedberg-Lee model and the quark parton model of Feynman has been resolved soon by the discovery of asymptotic freedom of quantum chromodynamics. 
It is a great honor for us to contribute to the special volume dedicated to the 95th anniversary of Professor Li.

\begin{figure}
\includegraphics[scale=1.5]{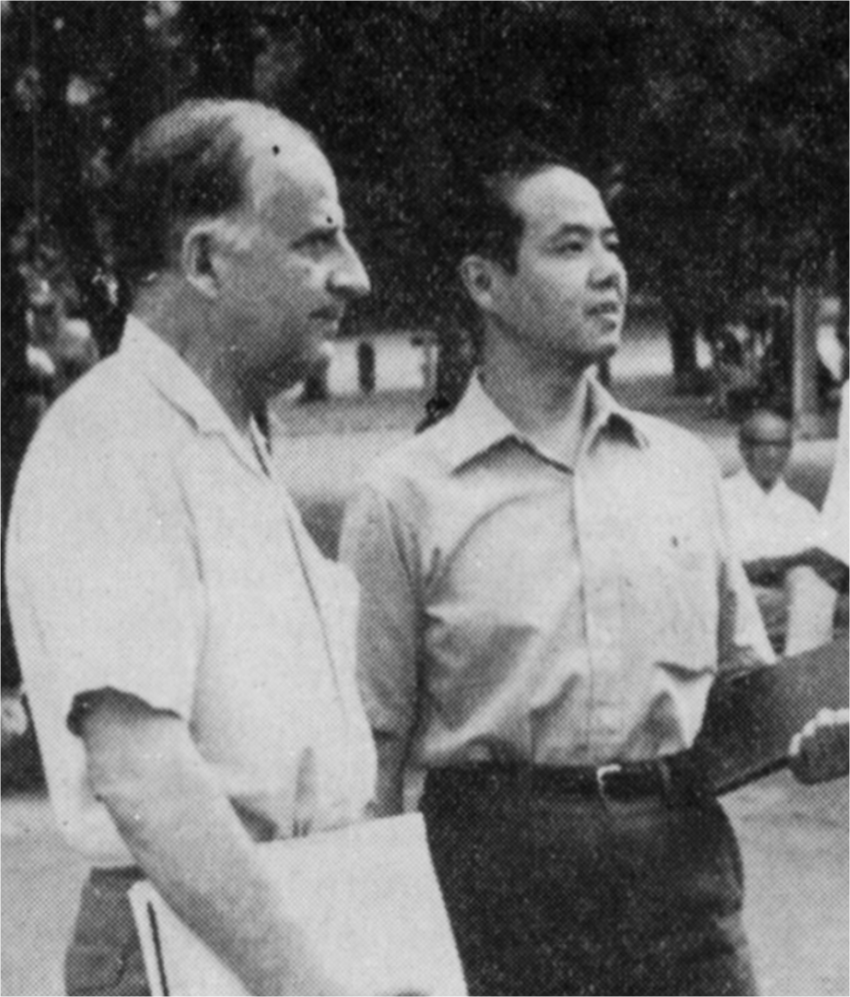}
\caption{ Zhengdao Li with Bruno Pontecorvo at the Balatonf\"ured Neutrino Conference in 1972}
\end{figure}

\section{Aims}

Axion electrodynamics, the Maxwell theory modified by  coupling the axion field to the electromagnetic topological density dates back to more than four decades of intense theoretical and phenomenological discussion \cite{sikivie}. 

 Following the unexpected discovery of the violation of parity conservation in weak $\beta$ decays suggested in 1956 by Lee and Yang, the violation of the combined CP symmetry was soon established in 1964. 
 With the advent of QCD, one might have also expected the breakdown of these symmetries in strong interactions. The coefficient of the allowed CP-violating term in the QCD Lagrangian is proportional to an angle which turned out to have an upper bound $10^{-10}$, derived from the observed upper bound on the electric dipole moment of neutron. This somewhat unnatural feature requires explanation. By far the most convincing interpretation has been proposed by Peccei and Quinn \cite{peccei77}, who introduced a pseudoscalar field, the axion replacing the CP-violating angle and suggesting a mechanism relaxing its expectation value to zero. Coherent oscillations of the axion field around this average offer an attractive alternative for the cosmologically evidenced dark matter. 

The axion field also couples  to electrodynamical degrees of freedom. Questions of modified light propagation through this medium are mostly in the focus of this research. The first studies  of axion propagation performed in quantum mechanical framework investigated photon--axion conversion via the Primakoff process \cite{maiani86,raffelt88}. 
These investigations were improved recently  using the formalism of quantum field theory \cite{arza19}, including propagation in curved space time \cite{capolupo19}. Plasmon--axion interactions receive similar attention \cite{mikheyev98,mendona20,caputo20,millar21}. 
Increasing accuracy in estimating the intensity of the solar axion flux \cite{jaeckel21} as well as axion effects arising in the strong magnetic field of neutron stars \cite{zhuravlev21} carry important discovery potential. A specific chapter of such investigations is represented by the detection techniques of axions, e.g., the theory of haloscopes \cite{millar17}.  
 
It is somewhat surprising that a sufficiently detailed general analysis of the issue of energy--momentum conservation in axion electrodynamics, a most important consequence of the space--time translation symmetry is mostly bypassed, despite some recent (partly controversial) publications \cite{nikitin12,tzompantzi21,tobar22,brevik22}. In particular, characterization of the energy transfer between axions and the electromagnetic field in the presence of a strong static magnetic field has a practical interest seeing that the haloscopes rely on the Primakoff process of photon--axion conversion. 
Although the invariance of the complete photon--axion system to space--time translations implies energy--momentum conservation, from the point of view of the electromagnetism of electrically charged objects, the existence of axions defines a specific dissipation channel.  Energy--momentum exchange through this channel might be significant for an indirect observation of axions.

The aim of the present paper is not to put forward some fundamentally new approach. It represents rather a pedagogically detailed discussion of the issue of energy--momentum conservation in axion electrodynamics, closely following the treatment of electrodynamical conservation laws and radiation backreaction presented in the classic textbook of \mbox{J.D. Jackson \cite{jackson}.}  Our complete analysis is classical! First, I  recall the modified Maxwell equations (presented by many authors and many times). Then, I proceed with extending the electromechanical energy conservation to the axion--electromechanical system. 
An  exact result is established {(see (\ref{mechanical-rate}) below)}, which is valid {beyond the linear approximation in} the strength of the axion--photon coupling. In the practically relevant case of extremely weak axion--photon coupling, it is demonstrated that the energy conservation naturally expressed in canonical variables is easily translated into an expression constructed of electric and magnetic field strengths.  
As an application of the general equations, I discuss the energy balance in presence of a spatially compact static magnetic field. An interesting point is the appearance of an effective axion equation arising from the elimination of the electromagnetic variables. This equation displays full-fledged radiation backreaction effect, damping axion oscillations.  The relaxation time is carefully estimated for the phenomenologically preferred axion mass range.

\section{Canonical Equations of Axion Electrodynamics and the Conservation of the Energy}
The Lagrangean density of the coupled axion ($a(x)$) + photon theory is the following:
\bea
&\displaystyle
L=\frac{1}{2}\left(\epsilon{\bf E}^2(x)-\frac{1}{\mu}{\bf B}^2(x)\right)-ga(x){\bf E}(x)\cdot{\bf B}(x)\nonumber\\
&\displaystyle
+\frac{1}{2}\left[\frac{1}{c^2}(\partial_ta(x))^2-(\partial_ia(x))^2-\frac{m_a^2c^2}{\hbar^2}a^2\right]-j_0A_0(x)+{\bf j}(x)\cdot{\bf A}(x),
\eea
with
\be
{\bf B}=\nabla\times{\bf A},\qquad {\bf E}=-\dot{\bf A}-\nabla A_0.
\ee

The homogeneous Maxwell equations follow from these relations:
\be
\nabla {\bf B}=0,\qquad \nabla\times{\bf E}=-\dot{\bf B}.
\ee

The canonical momenta are
\be
\Pi_a=\frac{1}{c^2}\dot a\equiv\frac{p_a}{c},\qquad \Pi_{A_0}=0,\qquad{\bf\Pi}_A=-\epsilon{\bf E}+ga{\bf B}\equiv -{\bf D}.
\ee

The Hamiltonian density is constructed by the usual Legendre transform:
\bea
&\displaystyle
{\cal H}=-\dot{\bf A}\cdot{\bf D}+\dot a\Pi_a-L\nonumber\\
&\displaystyle
=\frac{1}{2}\left[p_a^2+(\nabla a)^2+\frac{m_a^2c^2}{\hbar^2}a^2\right]+\frac{1}{\epsilon}{\bf D}\cdot({\bf D}+ga{\bf B})-\frac{1}{2\epsilon}({\bf D}+ga{\bf B})^2+\frac{1}{2\mu}{\bf B}^2\nonumber\\
&\displaystyle
+\frac{ga}{\epsilon}{\bf B}\cdot({\bf D}+ga{\bf B})+j_0A_0-{\bf j}\cdot{\bf A}\nonumber\\
&\displaystyle
=\frac{1}{2}\left[p_a^2+(\nabla a)^2+\frac{m_a^2c^2}{\hbar^2}a^2\right]
+\frac{1}{2}\left[\frac{1}{\epsilon}{\bf D}^2+\frac{1}{\mu}\left(1+\frac{\mu(ga)^2}{\epsilon}\right){\bf B}^2\right]+\frac{ga}{\epsilon} {\bf B}\cdot{\bf D}
\nonumber\\
&\displaystyle
+j_0A_0-{\bf j}\cdot{\bf A}.
\eea

The inhomogeneous equations are derived from the Hamiltonian formulation of the theory. 
Hamilton's equations can be written in the conventional form
with
\be
{\bf H}=\frac{1}{\mu^\prime}{\bf B}+\frac{1}{\epsilon}ga{\bf D},\qquad \frac{1}{\mu^\prime}=\frac{1}{\mu}\left(1+\frac{\mu(ga)^2}{\epsilon}\right)
\ee
\be
\nabla{\bf D}=j_0,\qquad \nabla\times{\bf H}={\bf j}+\dot{\bf D}, \qquad \ddot a-\Delta a+\frac{m_a^2c^2}{\hbar^2}a=-\frac{1}{\epsilon}g{\bf B}\cdot({\bf D}+ga{\bf B})=-g{\bf E}\cdot{\bf B}.
\label{hamilton-eq}
\ee

The rate of change in the energy in the axion sector is given as
\be
\frac{dE_a}{dt}=\int d^3x\dot a\left(\ddot a-\Delta a+\frac{m_a^2c^2}{\hbar^2}a\right)=-\int d^3xg\dot a{\bf E}\cdot{\bf B}\equiv \int d^3x{\bf j}_{axion}\cdot{\bf E}.
\label{axion-rate}
\ee

The extension of the mechanical energy conservation to axion electrodynamics starts by computing the rate of change in the mechanical energy through the work on an electromagnetic probe current ${\bf j}_e$ (which leads in the quasi-stationary approximation to Joule heat~generation):

\bea
&\displaystyle
\frac{dE_{mech}}{dt}=\int d^3x{\bf j}_e\cdot{\bf E}=\int d^3x\left[(\nabla\times{\bf H})\cdot{\bf E}-{\bf E}\cdot\dot{\bf D}\right]\nonumber\\
&\displaystyle
=\int d^3x\left[-\nabla({\bf E}\times{\bf H})-\dot{\bf B}\cdot{\bf H}-\dot{\bf D}\cdot{\bf E}\right]\nonumber\\
&\displaystyle
=\int d^3x\left[-\nabla({\bf E\times H})-\frac{1}{2\mu^\prime}\frac{d{\bf B}^2}{dt}-\frac{1}{2\epsilon}\frac{d{\bf D}^2}{dt}-\frac{ga}{\epsilon}\frac{d({\bf B}\cdot{\bf D})}{dt}\right],\nonumber\\
&\displaystyle
=-\int d^3x\left[\nabla({\bf E}\times{\bf H})+\frac{d}{dt}\frac{1}{2}\left(\frac{1}{\epsilon}{\bf D}^2+\frac{1}{\mu^\prime}{\bf B}^2\right)+\frac{d}{dt}\frac{g}{\epsilon}(a{\bf B}\cdot{\bf D})\right]
\nonumber\\
&\displaystyle
+\frac{g}{\epsilon}\int d^3x\dot a{\bf B}\cdot({\bf D}+ga{\bf B})\nonumber\\
&\displaystyle
=-\int d{\bf F}\cdot{\bf S}_{Poynting}-\frac{dE_{em+int}}{dt}+\int d^3x g\dot a{\bf B}\cdot{\bf E}.
\eea

One might be tempted to make exclusive use of the  "vacuum" field strengths: ${\bf E},{\bf B}$. Then, the mixed term disappears from the exact energy density of the electromagnetic field \cite{nikitin12}, and one finds
\be
\frac{dE_{mech}}{dt}=-\int d{\bf F}\cdot{\bf S}_{Poynting}- \frac{d}{dt}\frac{1}{2}\int d^3x\left(\frac{1}{\mu}{\bf B}^2+\epsilon{\bf E}^2\right)+\int d^3x g\dot a{\bf B}\cdot{\bf E}.
\label{mechanical-rate}
\ee

Adding any form of this rate to the rate of change in the energy in the axion sector, one recognizes that the power of the energy transfer between the two sectors is determined by
\be
W_{transfer}=\int d^3x g\dot a{\bf B}\cdot{\bf E}=-\int d^3x {\bf j}_{axion}\cdot{\bf E}
\label{transfer}
\ee
and the energy balance is exactly fulfilled: 
\bea
&\displaystyle
{\frac{d}{dt}\left[E_{mech}+\int d^3x\left(\frac{1}{\mu}{\bf B}^2+\epsilon{\bf E}^2\right)+
\int d^3x\frac{1}{2}\left(p_a^2+(\nabla a)^2+\frac{m_a^2c^2}{\hbar^2}a^2\right)\right]+\int d{\bf F}{\bf S}}\nonumber\\
&\displaystyle
=\frac{d}{dt}\left(E_{mech}+E_{e.m.}+E_{axion}\right)+\int d{\bf F}{\bf S}_{Poynting}=0.
\eea

{This expression displays a general feature of the energy--momentum tensor when expressed in terms of ${\bf E}$ and ${\bf B}$, namely the axion--photon interaction does not appear in it \cite{nikitin12} since the term (\ref{transfer}) responsible for the energy transfer between the two sectors mutually cancel between (\ref{axion-rate}) and (\ref{mechanical-rate}).}
One notes that with the present definitions of ${\bf D}$ and~${\bf H}$
\be
{\bf E}\times{\bf H}=\frac{1}{\epsilon\mu}{\bf D}\times{\bf B},\qquad  \epsilon{\bf E}^2+\mu{\bf H}^2=\frac{1}{\epsilon}{\bf D}^2+\frac{1}{\mu}{\bf B}^2+{\cal O}(g^2); 
\ee
therefore, the above expression with ${\cal O}(g)$ accuracy can be rewritten equally well in terms of the variables ${\bf E}$ and ${\bf H}$: 
\be
\frac{dE_{mech}}{dt}=\int d^3x\left[-\nabla({\bf E}\times{\bf H})-\frac{\partial}{\partial t}\left[\frac{1}{2}\left(\epsilon{\bf E}^2+\mu{\bf H}^2\right)-\mu ga{\bf E}\cdot{\bf H}\right]-{\bf j}_{axion}\cdot{\bf E}\right]+{\cal O}(g^2).
\ee


The Poynting vector (that is the integrand of the surface integral) is uniquely defined irrespective of the use of different field variables e.g., (${\bf E},{\bf H}$) or (${\bf D},{\bf B}$). {The apparent difference between the rate equation for ${\bf S}_{Abraham}={\bf D}\times{\bf B}$ and ${\bf S}_{Minkowski}={\bf E}\times {\bf H}$ observed in \cite{tobar22} is a consequence of restricting all their quantities to ${\cal O}(g)$ accuracy.}

\section{Energy Transfer from Axions in External Magnetic Field}

Consider a finite region of volume $V_H$ where a strong static magnetic induction field ${\bf B}_0({\bf x})$ is 
 present. One computes the electromagnetic field (${\bf E}_1,{\bf B}_1$) generated by the temporal variation of the axion field oscillation in this region, for one linearizes the dynamical Equation (\ref{hamilton-eq}) around the static magnetic field to find: 
\be
\frac{1}{\mu}\nabla\times{\bf B}_1=\dot{\bf D}=\epsilon(-\ddot{\bf A}_1-\nabla\dot A_{01})-g\dot a{\bf B}_0.
\ee 

Choosing the Lorentz gauge ($\nabla{\bf A}+\dot A_0=0$), an inhomogeneous wave equation follows: 
\be
\epsilon\mu\ddot{\bf A}_1-\Delta{\bf A}_1=-\mu g\dot a{\bf B}_0\equiv \mu {\bf j}_{axion}.
\ee

 With the help of the retarded solution to this equation 
\be
{\bf A}_1({\bf x},t)=\frac{\mu}{4\pi}\int d^3x^\prime\frac{1}{R}{\bf j}_{axion}({\bf x},t-R/c),\qquad R=|{\bf x}-{\bf x}^\prime| 
\label{axion-vector-potential}
\ee
we proceed to  calculate the rate of change in the axion energy due to the electromagnetically mediated self-action of the axion field. This idea is analogous to the computation of the radiation backreaction on the motion of electrically charged objects. 
Its interest has been emphasized recently in \cite{brevik22}, though the attention of that paper is focused on Joule heat production by the electric field produced by (\ref{axion-vector-potential}) in dielectrics in the quasi-stationary approximation.  

Our interest lies in estimating the size of the radiative backreaction on the axion field. The steps of the calculation might remind us the computation of the electromagnetic self-energy of the "Abraham-electron". The rate of change is given as
\be
\frac{dE_a}{dt}=\int d^3x {\bf j}_{axion}\cdot{\bf E}_1=\int d^3x {\bf j}_{axion}\cdot\left(-\dot{\bf A}_1-\nabla A_{01}\right).
\ee

The contribution from the second term vanishes assuming homogeneity of the axion field, which can be seen by performing partial integration and recognizing that $\nabla{\bf j}_{axion}=-g\dot a\nabla{\bf B}_0=0$.
Next, one substitutes (\ref{axion-vector-potential}) for the first term and arrives for the time-averaged energy loss at
\bea
&\displaystyle
\frac{dE_a}{dt}=-\frac{\mu}{16\pi}\int d^3x\int d^3x^\prime\frac{1}{R}~~~~~~~~~~~~~~~~~~~~~~~~~~~~~~~~~~~~~~~~~~~~~~~~~~~~~~~~~~~~~~\nonumber\\
&\displaystyle
\times\left[{\bf j}_{axion}({\bf x},t)\cdot\frac{\partial}{\partial t}{\bf j}_{axion}^*({\bf x}^\prime,t-R/c)
+{\bf j}_{axion}^*({\bf x},t)\cdot\frac{\partial}{\partial t}{\bf j}_{axion}({\bf x}^\prime,t-R/c)\right]\nonumber\\
&\displaystyle
=-\frac{\mu g^2}{16\pi}\int d^3x\int d^3x^\prime \frac{1}{R}{\bf B}_0({\bf x})\cdot{\bf B}_0({\bf x}^\prime)\left[\dot a(t)\ddot a^*(t-R/c)+\dot a(t)^*\ddot a(t-R/c)\right].
\label{retarded-energy-loss}
\eea

Assuming homogeneous harmonic oscillation of the axion field, $a(t)=a_0\exp(-i\omega_at)$, one arrives at
\be
-|a_0|^2\omega_a^3\frac{\mu g^2}{8\pi}\int d^3x\int d^3x^\prime \frac{\sin\left(\omega_aR/c\right)}{R}{\bf B}_0({\bf x})\cdot{\bf B}_0({\bf x}^\prime).
\label{exact-energy-loss-formula}
\ee

If in the volume where ${\bf B}_0\neq 0$ the retardation is small, e.g., $\omega_aR/c<<1$, then the sine can be approximated by the first term of its Taylor series:
\be
\frac{dE_a}{dt}\approx -|a_0|^2\omega_a^4\frac{\mu g^2}{8\pi c}\left(\int d^3x{\bf B}_0({\bf x})\right)^2. 
\ee

This result can also be obtained working with the starting expression without assuming the strict harmonic time dependence. In Equation~(\ref{retarded-energy-loss}),
  terms depending on the retarded moment can be expanded in Taylor series:
\be
\ddot a(t-R/c)=\sum_{n=0}^\infty \frac{1}{n\!}\left(-\frac{R}{c}\right)^n\frac{\partial^n}{\partial t^n}\ddot a(t).
\ee

One can average the power of the self-work of the axion field over the period of the nearly harmonic time dependence of $a(t)$. One promptly recognizes that the average of the $n=0$ term is zero. The leading nonzero contribution comes from the $n=1$ term. The time average is written after performing a partial integration as
\be
\frac{1}{T}\int_0^Tdt\frac{dE_a}{dt}=-\frac{\mu g^2}{8\pi c}\left(\int_{V_H}d^3x{\bf B}_0({\bf x})\right)^2\frac{1}{T}\int_0^T|\ddot a(t)|^2.
\ee

The relative magnitude of the higher terms is controlled by the combination $R_H\omega_a/c$, where $R_H$ is the size of the region ${\bf B}_0\neq 0$ and $\omega_a$ the characteristic frequency of the axion field. If the wavelength of the axion is much larger than this size, then these higher terms are negligible.
Below, for numerical estimates, we use the following quantities (which appear the most frequently when discussing haloscopes):
\be 
R_H=10 {\textrm {cm}},\qquad 10^{-16}{\textrm{GeV}}{/c^2}<m_a=\frac{\hbar\omega_a}{c^2}<10^{-14}{\textrm{GeV}}{/c^2},
\ee
using the following conversion from the MKSA to natural units:
\be
1{\textrm{cm}}=5.06\cdot 10^{13}{\textrm{GeV}}^{-1},
\ee 
one finds for this range of the axion mass
\be
0.05<\frac{\omega_a}{c}R_H<5.06.
\ee

This means that the expansion in retardation is legitimate for $m_a\leq 10^{-16}{\textrm{GeV}}$ but cannot be used in the higher mass range. Then, one has to return to (\ref{exact-energy-loss-formula}) and perform the symmetric double integral over the volume of the nonzero ${\bf B}_0$ region.

\section{Axion Oscillation in Presence of a Magnetic Field}

 The equation describing the homogeneous axion oscillation can be written after substituting the expression of ${\bf E}_1({\bf x},t)$:
\be
\frac{1}{c^2}\ddot a+\frac{m_a^2c^2}{\hbar^2}a=-\frac{\mu g^2}{4\pi}\int d^3x^\prime \frac{{\bf B}_0({\bf x}) 
\cdot {\bf B}_0({\bf x}^\prime)}{R}\ddot a\left(t-\frac{R}{c}\right).
\ee

For a semiquantitative estimate of the decay rate, we assume that the magnetic induction is homogeneous in the volume $V_H$ and average the right hand side to erase the apparent dependence on $\bf x$:
\be
\ddot a(t)+\frac{m_a^2c^4}{\hbar^2}a(t)=-\frac{\mu g^2}{2\pi}\left(\frac{1}{2}V_H{\bf B}_0^2\right)\frac{1}{V_H^2}\int_{V_H} d^3x\int_{V_H} d^3x^\prime\frac{1}{R}\ddot a\left(t-\frac{R}{c}\right).
\ee

First, we assume the expansion in retardation is justified. 
The leading contribution arises by neglecting retardation effects and is just proportional to $\ddot a$.
Putting together this piece with the original kinetic term, a new propagation velocity (smaller than the light velocity in a vacuum) can be defined after averaging this expression over the volume of the \mbox{${\bf B}_0\neq 0$ region:}
\be
\frac{1}{v_f^2}=\frac{1}{c^2}+\frac{\mu g^2}{4\pi}\frac{1}{V_H}\int d^3x\int d^3x^\prime\frac{1}{R}{\bf B}_0({\bf x}^\prime)\cdot{\bf B}_0({\bf x})
\ee

This would produce harmonic oscillation of the axion field with frequency \linebreak \mbox{$\omega_a=m_acv_f/\hbar$.} 

Radiative damping of the axion oscillations arises when going further in retardation effects. A force term $\sim \dddot a$ comes from the next ($n=1$) term of the expansion:
\be
\frac{\mu g^2}{4\pi c}\int d^3x^\prime {\bf B}_0({\bf x}^\prime)\cdot{\bf B}_0({\bf x})\dddot a(t)\approx \frac{\mu g^2}{4\pi c} V_H\left(\overline{{\bf B}_0}^{V_H}\right)^2\dddot a 
\ee

 This kind of self-force is completely analogous to the case of the "self-accelerating Abraham-electron", which leads to the following equation of motion for the axion amplitude in the long-wavelength limit:
\be
\ddot a+\frac{m_a^2c^2v_f^2}{\hbar^2}a=\frac{\mu g^2v_f^2}{4\pi c} V_H\left(\overline{{\bf B}_0}^{V_H}\right)^2\dddot a.
\label{damped-oscillator}
\ee

This equation coincides formally with the equation of the radiatively damped charged harmonic oscillator:
\be
\ddot x-\tau \dddot x=-\omega_0^2x,
\ee
with the correspondence
\be
\tau=\frac{\mu g^2v_f^2}{4\pi c} V_H\left(\overline{{\bf B}_0}^{V_H}\right)^2,\qquad \omega_0^2=\frac{m_a^2c^2v_f^2}{\hbar^2}.
\ee

The causal solution (excluding the axion amplitude runaway)  of (\ref{damped-oscillator}) can be constructed following Jackson's treatment, which gives with ${\cal O}(g^2)$ accuracy 
\be
a(t)=a_0e^{-\alpha t},  \qquad \alpha=\frac{\Gamma}{2}+i\omega_0, \qquad \Gamma=\tau\omega_0^2\approx \frac{\mu g^2c}{4\pi } V_H\left(\overline{{\bf B}_0}^{V_H}\right)^2\left(\frac{m_ac^2}{\hbar}\right)^2 .
\label{a-ansatz}
\ee

In natural ($\hbar=c=1$) units
\be
\Gamma=\frac{m_a^2}{2\pi}g^2U_B
\ee
where $U_B$ is the magnetic energy of the ${\bf B}_0\neq 0$ region. In addition to those listed at the end of the previous section, we choose 
\be
 m_a=10^{-16} {\textrm {GeV}}{/c^2}, \qquad |{\bf B}_0|=10{\textrm  {T}},\qquad g=10^{-12}{\textrm {GeV}}^{-1}.
\ee

One finds $U_B\approx 2.5\cdot 10^{14}{\textrm{GeV}}$, which provides for the radiative relaxation time of the coherent axion oscillations, an estimate comparable to the age of the Universe:
\be
\Gamma^{-1}\approx 1.7\cdot 10^{18}{\textrm{ s}}\approx 5.3\cdot 10^{10}{\textrm {yr}}.
\ee

For an estimate valid at the other edge of the mass range ($m_a=10^{-14}{\textrm{GeV}}{/c^2}$), we have to return to the equation describing the homogeneous axion oscillation without relying on the expansion in the retardation:
\be
\frac{1}{c^2}\ddot  a+\frac{m_a^2c^2}{\hbar^2}a=-\frac{\mu g^2}{4\pi}\int d^3x^\prime \frac{{\bf B}_0({\bf x})\cdot{\bf B}_0({\bf x}^\prime)}{R}\ddot a\left(t-\frac{R}{c}\right).
\ee

For a semiquantitative estimate of the decay rate, we assume that the magnetic induction is homogeneous in the volume $V_H$ and average the right-hand side to suppress the apparent dependence on $\bf x$. After substituting the approximate solution in the form introduced in (\ref{a-ansatz}), one has 
\be
\alpha^2 a(t)+\frac{m_a^2c^4}{\hbar^2}a(t)=-\frac{\mu g^2\alpha^2}{2\pi}\left(\frac{1}{2}V_H{\bf B}_0^2\right)\frac{1}{V_H^2}\int_{V_H} d^3x\int_{V_H} d^3x^\prime\frac{e^{\alpha R/c}}{R}\cdot a(t).
\ee

From this, the dispersion relation is extracted. Its real part can be used to find $\omega_0$,  which is slightly shifted from the frequency $\omega_a$. Still, $\omega_0R_H/c$ is larger than unity, therefore one cannot expand into Taylor series the exponential $\exp(-i\omega_0R/c)$. However, one can assume $\Gamma<<\omega_0$ and linearize $\exp(\Gamma R/2c)$ in $\Gamma$.
Then, the imaginary part of the resulting approximate equation can be used for the determination of the decay rate:
\be
\omega_0\Gamma=-\frac{\mu g^2c^2}{2\pi}\left(\frac{1}{2}V_H{\bf B}_0^2\right) \frac{1}{V_H^2}\int_{V_H} d^3x\int_{V_H} d^3x^\prime\left[-\omega_0^2\frac{\sin(\omega_0 R/c)}{R}\left(1+\frac{R}{2c}\right)+\omega_0\Gamma\frac{\cos(\omega_0 R)}{R}\right].
\ee

The integral of the quickly oscillating integrand $\sin(\omega_0 R/c)$ is very close to  zero. Rearranging the terms one finds
\bea
&\displaystyle
\Gamma\left[1+\frac{\mu g^2c^2}{2\pi}\left(\frac{1}{2}V_H{\bf B}_0^2\right) \frac{1}{V_H^2}\int_{V_H} d^3x\int_{V_H} d^3x^\prime\frac{\cos(\omega_0R/c)}{R}\right]\nonumber\\
&\displaystyle
=\frac{\omega_0\mu g^2c^2}{2\pi}\left(\frac{1}{2}V_H{\bf B}_0^2\right) \frac{1}{V_H^2}\int_{V_H} d^3x\int_{V_H} d^3x^\prime\frac{\sin(\omega_0R/c)}{R}.
\label{radiative-width}
\eea

The quantities appearing in this equation are related to the real and imaginary part of the integral:
\be
I(k_0,R_H)=\frac{1}{V_H^2}\int_{V_H} d^3x\int_{V_H} d^3x^\prime\frac{e^{ik_0R}}{4\pi R},\qquad R=|{\bf x}-{\bf x}^\prime|,\quad k_0=\frac{\omega_0}{c}.
\ee

Exploiting the expansion of the integrand in terms of spherical harmonics
\be
\frac{e^{ik_0R}}{4\pi R}=ik_0\sum_lj_l(k_0r_<)h_l^{(1)}(k_0r_>)\sum_{m=-l}^lY^*_{lm}(\Omega)Y_{lm}(\Omega^\prime)
\ee
the angular integrals can be performed. We choose a sphere of radius $R_H $ since it is very convenient for a fully analytic treatment. In addition, exploiting the symmetry of the range of integration one finds
\be
I(k_0,R_H)=\frac{8\pi ik_0}{V_H^2}\int_0^{R_H}r_>^2dr_>\int_0^{r_>}r_<^2dr_<j_0(k_0r_<)h_0^{(1)}(k_0r_>).
\ee

Introducing the notation $X_H=k_0R_H$, one easily finds
\bea
&\displaystyle
{\textrm{Im}}I(k_0,R_H)=\frac{9k_0}{4\pi X_H^6}\left[X_H^2+1+(X_H-1)\cos(2X_H)-2X_H\sin(2X_H)\right]\nonumber\\
&\displaystyle
{\textrm{Re}}I(k_0,R_H)=\frac{9k_0}{4\pi X_H^6}\left[-\frac{2}{3}X_H^3-(X_H^2-1)\sin(2X_H)-2X_H\cos(2X_H)\right].
\eea

With the actual parameters, the second term in the bracket on the left-hand side of (\ref{radiative-width}) is negligible relative to the unity. Second, on the right-hand side, the quantity multiplying $\omega_0$ is very small, ${\cal O}(10^{-27})$, which confirms the linearization in $\Gamma$. Due to the larger axion mass, a somewhat faster decay rate is obtained:
\be
\Gamma^{-1}\approx 5 \cdot 10^9{\textrm{yr}}.
\ee

This is still comparable to the age of the Universe.

\section{Momentum Conservation  with Axions}

Again, one starts with the rate of change in the mechanical momentum of the electrically charged medium:
\bea
&\displaystyle
\frac{d{\bf P}_{mech}}{dt}=\int d^3x(j_0{\bf E}+{\bf j}\times{\bf B})=-\frac{d}{dt}\int d^3x({\bf D}\times {\bf B})\nonumber\\
&\displaystyle
+\int d^3x\left[{\bf E}(\nabla{\bf D})+(\nabla\times {\bf H})\times{\bf B}-{\bf D}\times(\nabla\times{\bf E})+{\bf H}(\nabla{\bf B})\right].
\label{mechanical-force}
\eea

Changing from (${\bf D},{\bf B})$) to the (${\bf E},{\bf H}$) pair of variables, one notes the exact relation
\be
{\bf D}\times {\bf B}=(\epsilon{\bf E}-ga{\bf B})\times{\bf B}=\epsilon{\bf E}\times(\mu{\bf H}-\mu ga{\bf E})=\epsilon\mu{\bf E}\times{\bf H},
\ee
which gives an unchanged relation of the electromagnetic momentum density to the Poynting vector.

The {second volume integral} of the {second} expression of (\ref{mechanical-force}) is the simplest when using the pair of variables (${\bf E},{\bf B}$). Then, {$i$-th component of its integrand reads (summation is understood over repeated indices):}
\bea
&\displaystyle
{E_i(\nabla{\bf D})+[(\nabla\times {\bf H})\times{\bf B}]_i-[{\bf D}\times(\nabla\times{\bf E})]_i+ H_i(\nabla{\bf B})}\nonumber\\
&\displaystyle 
=E_i\partial_jD_j+B_j\partial_jH_i-B_j\partial_iH_j+D_j\partial_jE_i-D_j\partial_iE_j+H_i\partial_jB_j\nonumber\\
&\displaystyle
=\epsilon\left(\partial_j(E_iE_j)-E_j\partial_iE_j\right)-g\partial_j(E_iB_ja)+gaB_j\partial_iE_j\nonumber\\
&\displaystyle
+\frac{1}{\mu}\left(\partial_j(B_iB_j)-B_j\partial_iB_j\right)+g\partial_j(E_iB_ja)-gB_j\partial_i(aE_j)
\eea

By the cancellations between the last two lines one arrives at 
\bea
&\displaystyle
\frac{d{\bf P}_{mech,i}}{dt}=-\epsilon\mu\frac{d}{dt}\int d^3x({\bf E}\times {\bf H})_i-\int d^3x\partial_j\Pi_{ij}-\int d^3x\partial_ia({\bf E}\cdot{\bf B}),\nonumber\\
&\displaystyle
\Pi_{ij}=\left[\frac{1}{2}\left(\epsilon{\bf E}^2+\frac{1}{\mu}{\bf B}^2\right)\delta_{ij}-\epsilon E_iE_j-\frac{1}{\mu} B_iB_j\right].
\eea

Momentum leaks to the axion sector through inhomogeneity of the axion field. In the long-wavelength limit there is no recoil from the axion sector. This is consistent with the vanishing Lorentz self-force, when the spatial variation of the axion is neglected, e.g., ${\bf j}_a\times {\bf B}=-g\dot a{\bf B}\times{\bf B}=0$. A spatial gradient of the axion field splits the electromagnetic eigenmodes between parallel uncharged material plates, though the summary contribution to the Casimir force
\cite{brevik-casimir}.

\section{Summary}
A pedagogical derivation of the classical laws of energy--momentum conservation in Maxwell's electrodynamics completed by the interaction of the axion field with the topological charge density of the electromagnetic fields was presented. This derivation allows the unique identification of the term responsible for the energy transfer between the electromagnetic and the axion sectors of the theory. Its power density ${\bf j}_{axion}{\bf E}$ is completely analogous to the energy dissipation power density of electrically charged objects. It allowed the computation of the backreaction of the  electromagnetic radiation emitted by the axion field oscillations on these oscillations themselves. An explicit example is presented for the case of radiation in the presence of a strong static magnetic field in a finite macroscopic region. The characteristic damping time, estimated here for a first time, is comparable to the age of the Universe for typical values of axion masses and haloscope size.

\section*{Acknowledgements} This research received support of the Hungarian National Research, Development and Innovation Fund under Project No. K123815. Discussions with Tam\'as Bir\'o and Gergely Fej\H os are gratefully acknowledged.

\end{document}